\documentclass[runningheads]{llncs}
\usepackage[T1]{fontenc}
\usepackage{pgfplotstable}
\usepackage{booktabs}
\usepackage{tabularx}
\usepackage{graphicx}
\usepackage{multirow}
\usepackage{caption}
\usepackage{booktabs}
\usepackage{tikz,hyperref}
\usepackage{subcaption}
\usepackage{adjustbox}
\usepackage{rotating}
\usepackage{listings}
\usepackage{pgfplots}
\pgfplotsset{compat=1.18}
\usepackage{pifont}
\usepackage{todonotes}

\usepackage{orcidlink}

\begin{document}

\title{A Multi-Dataset Evaluation of Models for Automated Vulnerability Repair}
\author{Zanis Ali Khan\orcidlink{0000-0002-3935-2148} \and
Aayush Garg\orcidlink{0000-0002-2507-8846} \and
Qiang Tang\orcidlink{0000-0002-6153-4255}}
\authorrunning{Zanis Ali Khan, Aayush Garg, and Qiang Tang}
\institute{Luxembourg Institute of Science and Technology (LIST), Luxembourg
\email{\{zanis-ali.khan, aayush.garg, qiang.tang\}@list.lu}}

\maketitle       
\begin{abstract}
Software vulnerabilities pose significant security threats, requiring effective mitigation. While Automated Program Repair (APR) has advanced in fixing general bugs, vulnerability patching—a security-critical aspect of APR—remains underexplored. This study investigates pre-trained language models, CodeBERT and CodeT5, for automated vulnerability patching across six datasets and four languages. We evaluate their accuracy and generalization to unknown vulnerabilities. Results show that while both models face challenges with fragmented or sparse context, CodeBERT performs comparatively better in such scenarios, whereas CodeT5 excels in capturing complex vulnerability patterns. CodeT5 also demonstrates superior scalability. Furthermore, we test fine-tuned models on both in-distribution (trained) and out-of-distribution (unseen) datasets. While fine-tuning improves in-distribution performance, models struggle to generalize to unseen data, highlighting challenges in robust vulnerability detection. This study benchmarks model performance, identifies limitations in generalization, and provides actionable insights to advance automated vulnerability patching for real-world security applications.

\end{abstract}

\section{Introduction}

Software vulnerabilities remain a constant threat to contemporary software systems, 
leaving them susceptible to exploitation by malicious actors. These vulnerabilities, 
which include problems like injection flaws and memory management errors, can result in 
unauthorized access, data breaches, and service interruptions~\cite{OkutanMMKSGS23}. 
Addressing these issues is essential to ensure the reliability and security of software 
systems~\cite{AlbaneseCJ14}. However, the manual effort required to detect and fix 
these vulnerabilities is time-consuming, prone to errors, and struggles to match 
the growing complexity and scale of today's software ecosystems~\cite{KhanP18}.

Automated Program Repair (APR) has gained traction as a promising approach to tackle 
this issue, employing computational methods to autonomously generate fixes for 
software bugs~\cite{BuiPVMS24}. While APR has achieved notable progress in 
addressing general software defects, the specialized area of vulnerability-focused 
program repair—which deals with the unique challenges of security vulnerabilities—remains 
underdeveloped. Unlike general-purpose bug fixes, patches for vulnerabilities often require addressing 
the flaw in a way that is not only functionally correct but also generalizable across variations of 
the vulnerability~\cite{deFiteroDominguezGGM24}. This makes vulnerability patching a more nuanced 
subset of Automated Program Repair (APR), where the ability to generate broadly applicable fixes 
becomes especially important.

Existing techniques in vulnerability-focused Automated Program Repair (APR) 
predominantly depend on either static analysis tools or traditional machine 
learning models trained on specific vulnerability patterns. Although these 
approaches have demonstrated potential in identifying vulnerabilities, their 
capacity to generate meaningful and effective patches remains limited. For 
instance, static analysis tools are highly effective at detecting vulnerabilities 
but often struggle to produce practical fixes~\cite{PiskachevBB23}. Similarly, 
conventional machine learning models are hindered by their dependence on 
restricted datasets~\cite{abs-2012-11701}, which limits their generalizability 
and effectiveness across a wide range of programming languages and vulnerability types~\cite{RisseB24}.

Recent advancements in deep-learning have paved the way for automated vulnerability 
patching, particularly with the emergence of pre-trained language models tailored for 
code. Models like CodeBERT\cite{feng2020codebert} and CodeT5\cite{wang2021codet5} utilize 
large-scale code corpora to capture both syntactic and semantic structures, 
facilitating tasks such as code generation, summarization, and translation~\cite{GargDPT24}. 
Their ability to discern patterns from extensive datasets makes them a promising tool for 
vulnerability-focused program repair. However, the practical application of these models 
remains challenging. Due to substantial differences in syntax, semantics, and vulnerability 
characteristics across programming languages, existing pre-trained models, which are often 
designed for monolingual or domain-specific tasks, may struggle with 
generalization~\cite{Dang2024}. Evaluating their performance across diverse languages is 
therefore a crucial yet underexplored area of research~\cite{GharibiSF24}.

This paper systematically evaluates the performance of pre-trained language models in vulnerability-focused program repair, specifically analyzing CodeBERT and CodeT5 in generating patches for known vulnerabilities across six datasets covering four programming languages. We assess their effectiveness using \emph{CodeBLEU} and \emph{CrystalBLEU} scores and explore their generalizability by evaluating performance on both in-distribution and out-of-distribution datasets, providing insights into their strengths and limitations.

Our results show that while both models excel in generating vulnerability patches, they exhibit distinct limitations. CodeT5 generally outperforms CodeBERT in accuracy, especially on datasets with complex vulnerability patterns. However, both models struggle with fragmented contexts and sparse data, which limits their ability to produce correct fixes in such settings. Additionally, while fine-tuning improves performance on in-distribution datasets, both models face challenges in generalizing to out-of-distribution datasets, highlighting limitations in detecting and patching vulnerabilities in unseen scenarios.


Hence, our contributions in this paper are threefold:
\begin{itemize}
    \item We provide an evaluation of CodeBERT and CodeT5 for vulnerability-focused program repair, covering a diverse set of 6 datasets across multiple programming languages.
    \item We establish benchmarks for model performance in generating vulnerability patches, serving as a foundation for evaluating pre-trained models in dataset-driven vulnerability patching scenarios.
    \item We identify key limitations in model generalization, particularly the challenges of fine-tuning and performance on out-of-distribution datasets.
\end{itemize}

\section{Related Work}
Software vulnerabilities refer to security gaps or defects within code that can be leveraged by malicious actors to compromise systems~\cite{ogata2018vetting}. One notable example is the buffer overflow vulnerability, which arises when a program tries to write more data into a buffer than it can hold, leading to the overflow spilling into neighboring memory areas. This can allow attackers to inject and execute harmful code~\cite{GargPKT24}. As these vulnerabilities grow more complex, they pose substantial obstacles to developing and deploying robust countermeasures.  

While vulnerability detection has been extensively studied, significantly less attention has been given to generating patches. Traditional static analysis tools have long been used for detection, but their reliance on predefined rules often makes it difficult to identify complex patterns~\cite{static2017andrei}. In contrast, AI-driven methods have gained traction for their ability to process vast codebases and uncover intricate security flaws. Models like \emph{CodeBERT}\cite{codebert2020} and \emph{GraphCodeBERT}\cite{GuoRLFT0ZDSFTDC21} have proven effective in analyzing source code, contributing to advancements in vulnerability detection and assessment~\cite{abs-2303-04247}. Additionally, large language models (LLMs) such as OpenAI’s GPT-4, Meta AI’s Llama2, and Mistral AI’s Mistral have demonstrated strong adaptability in tackling vulnerability detection tasks~\cite{guo2023empirical}.

Conversely, creating effective patches continues to be a significant challenge. The majority of research on automated patch generation is centered on fixing general code defects rather than targeting vulnerabilities directly. The subsequent sections will explore methodologies within this broader context.

\subsection{Traditional Approaches to Code Repair}
Automated code repair traditionally falls into two categories: heuristic-based and constraint-based~\cite{goues2019automated}. Heuristic methods search for patches that pass all tests, often using transformation schemas for efficiency~\cite{liu2019tbar}. Approaches like GenProg~\cite{Le2012genprog} and PAR~\cite{Kim2013automatic} leverage genetic programming, while others use random or deterministic strategies to refine the search.

Constraint-based methods employ symbolic execution~\cite{Cadar2008klee} to guide patch generation by exploring multiple execution paths. Tools such as SemFix~\cite{semfix2013} and Angelix~\cite{angelix2016} derive repair constraints, while techniques like Nopol~\cite{Nopol2017} target specific cases, such as repairing conditional expressions.



\subsection{ML-Based Code Repair}
Machine learning has emerged as a key technique for automating code repair, generating patches for software vulnerabilities and bugs. Early efforts relied on Neural Machine Translation (NMT) with encoder-decoder architectures, such as SequenceR~\cite{SequenceR2021} and CODIT~\cite{Chakraborty2022edit}, which used attention mechanisms to prioritize critical regions during decoding.

More recently, transformer-based models have excelled at capturing long-range dependencies and nuanced context, leveraging attention to focus on relevant code segments. Ding \emph{et al.}~\cite{Ding2020metaphor} highlighted their transformative potential, paving the way for broader adoption in program repair.

Further expanding these approaches, large language models (LLMs) such as CodeBERT~\cite{feng2020codebert} and CodeT5~\cite{wang2021codet5} have shown promise for code-related tasks, benefiting from pretraining on large code corpora. While prior work has explored their capabilities in general code generation and repair, their effectiveness for vulnerability-specific patching remains underexplored. This motivates our evaluation of both models in this context.

Nevertheless, patching vulnerabilities is distinct from fixing general bugs. It requires highly contextual, security-focused modifications and robust generalization across complex scenarios. Current solutions emphasize fine-tuning LLMs and advancing techniques to enhance adaptability for various datasets and security-specific demands.

\section{Methodology}\label{sec:methodology}
In this section, we outline our experimental workflow, from dataset preparation and preprocessing to splitting the data for training and testing, followed by model selection and fine-tuning strategies.

\subsection{Dataset Preparation and Pre-processing}\label{subsec:pre-processing}

For this study, we collected six publicly available datasets containing code samples with known 
vulnerabilities and their corresponding patches. These datasets comprises of multiple programming 
languages, including Go, Java, PHP, and C, ensuring diverse code structures and vulnerability 
patterns. The inclusion of diverse datasets allowed us to evaluate the models' ability to generalize across varied programming contexts. Details about these datasets, including their references are 
provided in Section~\ref{subsec:datasets}, offering a comprehensive overview of their sources. This 
diversity in datasets not only enhances the robustness of our evaluation but also reflects real-world 
scenarios where vulnerabilities span multiple languages and coding paradigms.

We preprocessed the raw datasets to standardize their structure and enhance model compatibility. 
Given the noise in real-world vulnerability datasets~\cite{abs-2012-11701,KhanSBB24}, our 
preprocessing aimed to reduce inconsistencies and improve data quality, as emphasized in 
studies on noisy datasets~\cite{Khan0BB22,GargDJCPT22}. By ensuring uniformity, we created 
a robust foundation for reliable model training and evaluation. These steps were critical 
for noise reduction and dataset preparation.

\begin{enumerate}
    \item[i.] \textbf{Token Length Filtering.} Code exceeding \emph{512} tokens was truncated/excluded due to model limits.
    \item[ii.] \textbf{Comment Removal.} Language-specific regex removed comments, focusing on functional code.
    \item[iii.] \textbf{Normalization.} Fixed formatting inconsistencies (whitespace, line breaks) for uniform datasets.
\end{enumerate}

\begin{table}[htbp] 
\centering 
\small
\caption{Datasets}
\vspace{0.5em}
\begin{filecontents*}{data/statistics_ease_paper.csv}
dataset,totalRowsInitial,totalRowsRemaining,impactedTokenization,impactedComments,impactedNormalization,totalRowsRemaining,testInstances
Go,1472,921,551,357,0,921,138
PHP,6696,6360,335,4923,1,6360,954
MegaVul_C_2023 ,17975,14526,3147,0,302,14526,2179
MegaVul_C_2024,17975,14526,3147,0,302,14526,2179
Vul4J,1790,1790,3,0,0,1790,269
CodeParrot,69420,50000,19420,1505,0,50000,7500
\end{filecontents*}
\pgfplotstabletypeset[
    col sep=comma,
    fixed,
    empty cells with={-},
    string replace={---}{\textemdash},
     string replace={_}{\textunderscore}, 
    every head row/.style={
        before row={\toprule},
        after row=\midrule,
    },
    every last row/.style={after row=\bottomrule},
    columns={dataset,totalRowsInitial,impactedTokenization,impactedComments,impactedNormalization,totalRowsRemaining}, 
    columns/dataset/.style={column type=l, column name=Dataset, string type, postproc cell content/.code={%
        \pgfplotsutilstrreplace{_}{\_}{##1}%
        \pgfkeyslet{/pgfplots/table/@cell content}\pgfplotsretval
    },},
    columns/totalRowsInitial/.style={column type=r, column name= $I_{rows}$, precision=4},
    columns/impactedTokenization/.style={column type=r, column name=$R_{tok.}$, precision=4},
    columns/impactedComments/.style={column type=r, column name=$R_{comm.}$, precision=4},
    columns/impactedNormalization/.style={column type=r, column name=$R_{norm.}$, precision=4},columns/totalRowsRemaining/.style={column type=r, column name=$T_{rows}$, precision=4},
]{data/statistics_ease_paper.csv} 

\label{table:datasets}
\vspace{-1.5em}
\end{table}

\subsection{Training and Testing Split}\label{subsec:test-train-split}

The datasets were partitioned into \emph{85\%} for training and \emph{15\%} for testing, a widely adopted ratio that provides a robust balance between model learning and evaluation~\cite{zhao2020maritime}. 
This split ensures sufficient data for effective training while reserving enough samples to yield meaningful test results. To avoid data leakage and maintain the integrity of the evaluation, all overlapping 
or duplicate instances were excluded.

\subsection{Model Selection and Fine-Tuning}
We utilized and fine-tuned \emph{CodeBERT} and \emph{CodeT5} for vulnerability patching, leveraging their strengths in code understanding and generation. CodeBERT, tailored for programming tasks, adapted to detect vulnerabilities and their fixes, while CodeT5, optimized for code generation, improved handling diverse code structures. Despite alternatives like \emph{TFix}, these models were chosen for their versatility, robustness, and real-world applicability.

\section{Experimental Setup}
In this section, we detail the computational environment and methodologies used for training and evaluating our models on a range of vulnerable code scenarios. All experiments were conducted on a High-Performance Computing (HPC) cluster with nodes featuring \textit{2.20GHz Intel Xeon Silver 4210} processors and \textit{NVIDIA Tesla V100-PCIE-32GB} GPUs. Model training and evaluation were performed using the \textit{PyTorch 2.0.1} framework with \textit{CUDA 12} compatibility.

\subsection{Datasets}\label{subsec:datasets}
To address the research questions outlined in Section~\ref{sec:results}, we leveraged 
publicly available datasets that contain comprehensive collections of vulnerable source 
code along with their corresponding fixed versions, which served as our ground 
truth. Specifically, we utilized six datasets, including 
Go and PHP~\footnote{Go and PHP--\url{https://doi.org/10.5281/zenodo.13870382}}, 
MegaVul\_C\_2023, and MegaVul\_C\_2024~\footnote{MegaVul\_C\_2023, and MegaVul\_C\_2024--\url{https://github.com/Icyrockton/MegaVul}}~\cite{10.1145/3643991.3644886}, Vul4J\footnote{Vul4J--\url{https://github.com/tuhh-softsec/vul4j}}\cite{vul4j2022}, and also 
CodeParrot~\footnote{\url{https://huggingface.co/datasets/codeparrot/github-code-clean}}. These datasets encompass a variety of programming languages, including C, Java, 
Go, and PHP, offering a well-rounded foundation for evaluation. Prior to their use, 
we implemented preprocessing steps as outlined in Section~\ref{subsec:pre-processing}

\textbf{Table~\ref{table:datasets}} reports on the size of our datasets, in terms of the number of rows 
(\(I_{rows}\)), rows affected by tokenization ($R_{tok.}$), rows affected by 
comment removal ($R_{comm.}$), rows affected by normalization ($R_{norm.}$), 
and the total number of rows remaining after pre-processing ($T_{rows.}$). 

\begin{table}[t] 
\centering 
\small
\caption{Accuracy Scores}
\vspace{0.5em}
\begin{filecontents*}{data/summary_RQ1_RQ2_ease_paper.csv}
Dataset,CodeBLEUBERT,CodeBLEUT5,CrystalBLEUBERT,CrystalBLEUT5,trainTimeBERT,trainTimeT5,testTimeBERT,testTimeT5,totalTimeBERT,totalTimeT5,AvgInferenceTimeBERT,AvgInferenceTimeT5,testInstancesBERT,testInstancesT5
Go,0.76409358,0.649903524,0.655735471,0.52641215,1308.211,2353.508,339.45,173.429,1647.661,2526.937,2.442086331,1.247690647,139,139
PHP,0.735140199,0.692412555,0.462377324,0.372659485,8747.032,16118.246,1405.674,1219.889,10152.706,17338.135,1.47345283,1.278709644,954,954
MegaVul_C_2023,0.839612819,0.854883959,0.789343064,0.813103693,20323.176,37135.945,4429.201,2653.073,24752.377,39789.018,2.032675998,1.217564479,2179,2179
MegaVul_C_2024,0.839544128,0.85492269,0.789343064,0.813103693,20247.623,36976.005,4422.296,3322.019,24669.919,40298.024,2.029507113,1.524561267,2179,2179
Vul4J,0.373723613,0.937347328,0.122904125,0.898485089,2488.586,4508.079,423.686,217.825,2912.272,4725.904,1.575040892,0.809758364,269,269
CodeParrot,0.996957727,0.997266102,0.959520402,0.960339439,67685.864,125451.923,6160.096,5937.041,73845.96,131388.964,0.821346133,0.791605467,7500,7500
\end{filecontents*}
\pgfplotstabletypeset[
    col sep=comma,
    fixed,
    empty cells with={-},
    string replace={---}{\textemdash},
    string replace={_}{\textunderscore},
    every head row/.style={
        before row={
	   		\toprule
			\multicolumn{1}{c}{} & 
			\multicolumn{2}{c}{CodeBLEU} &
			\multicolumn{2}{c}{CrystalBLEU} 
			\\
			\cmidrule(r){2-3}
			\cmidrule(r){4-5}
		},
        after row=\midrule,
    },
    every last row/.style={after row=\bottomrule},
    columns={Dataset,CodeBLEUBERT,CodeBLEUT5,CrystalBLEUBERT,CrystalBLEUT5}, 
    columns/Dataset/.style={column type=l, string type, postproc cell content/.code={%
        \pgfplotsutilstrreplace{_}{\_}{##1}%
        \pgfkeyslet{/pgfplots/table/@cell content}\pgfplotsretval
    }},
    columns/CodeBLEUBERT/.style={column type=r,column name=$CodeBERT$, precision=4},
    columns/CodeBLEUT5/.style={column type=r, column name=$CodeT5$, precision=4},
    columns/CrystalBLEUBERT/.style={column type=r,column name=$CodeBERT$, precision=4},
    columns/CrystalBLEUT5/.style={column type=r, column name=$CodeT5$, precision=4},
]{data/summary_RQ1_RQ2_ease_paper.csv}

\label{table:rq1_results}
\vspace{-1.5em}
\end{table}

\subsection{DL Models}\label{subsec:models}
For vulnerability patching, we employed \emph{CodeBERT}\cite{feng2020codebert} and \emph{CodeT5}\cite{wang2021codet5}, widely used for code analysis and vulnerability detection due to their strong performance in handling code semantics and structure.

\emph{CodeBERT}\cite{feng2020codebert} bridges programming and natural languages, enhancing tasks like code completion, summarization, and vulnerability detection. Built on a transformer architecture, it captures syntactic and semantic relationships from code–language pairs, enabling precise vulnerability identification and remediation at scale.

\emph{CodeT5}\cite{wang2021codet5} is a T5-based model for code generation and understanding, excelling in vulnerability detection and patching. It generates context-aware patches, preserves code intent, and supports multiple languages. Pre-trained on extensive programming data, it performs well on benchmarks, improving software security and code quality. It also preserves semantics in decompilation, advancing vulnerability repair frameworks~\cite{WuJPLD0BS23}.




\subsection{Evaluation Metrics}\label{subsec:metrics}

We evaluated the LLMs using \emph{CrystalBLEU}\cite{eghbali2022crystalbleu} and 
\emph{CodeBLEU}\cite{ren2020codebleu}. CrystalBLEU refines BLEU~\cite{papineni2002bleu} 
by addressing n-gram limitations in programming languages, focusing on trivially shared 
n-grams for better code evaluation. CodeBLEU enhances BLEU by combining n-gram matching 
with AST-based structures and semantic data flow, making it ideal for assessing code 
quality. Together, these metrics provide accurate evaluations by considering both syntactic 
and semantic aspects of generated code.

\begin{figure}[htbp]
    \centering
    \begin{subfigure}[b]{\textwidth}
        \includegraphics[width=\textwidth]{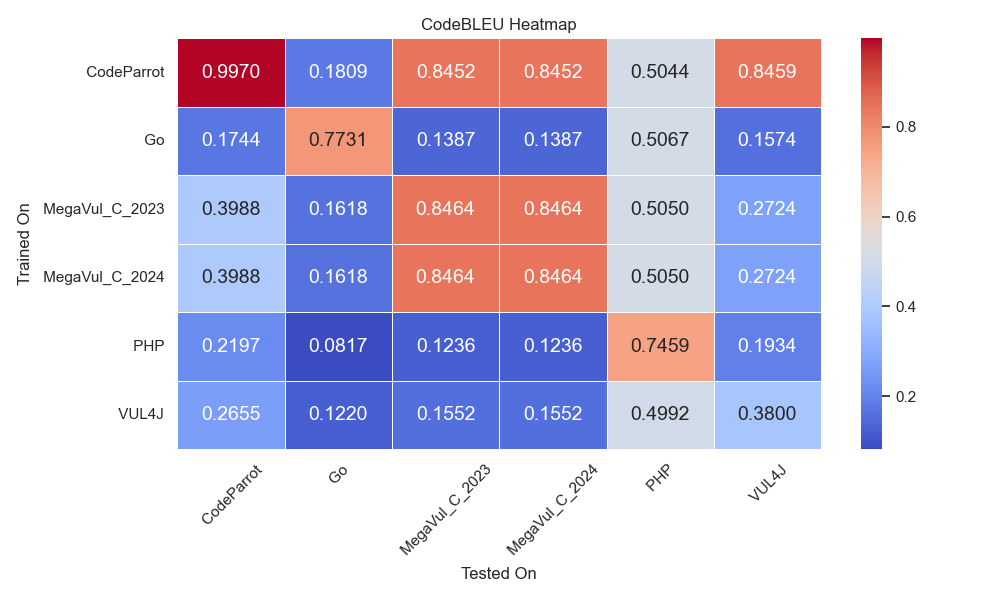}
        \caption{CodeBLEU for CodeBERT}
        \label{fig:codebert_codebleu}
    \end{subfigure}
    \hspace{0.02\textwidth}
    \begin{subfigure}[b]{\textwidth}
        \includegraphics[width=\textwidth]{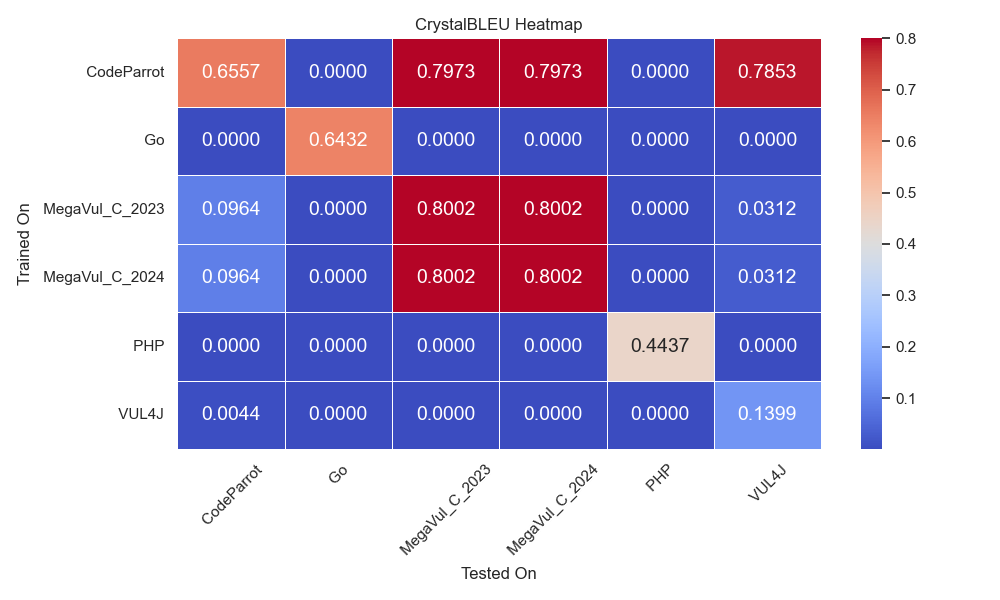}
        \caption{CrystalBLEU for CodeBERT}
        \label{fig:codebert_crystalbleu}
    \end{subfigure}
    \captionsetup{justification=centering} 
    \caption{Heatmaps for CodeBERT}
    \label{fig1:codebert}
\vspace{-1.5em}
\end{figure}


\begin{figure}[htbp]
    \centering
    \begin{subfigure}[b]{\textwidth}
        \includegraphics[width=\textwidth]{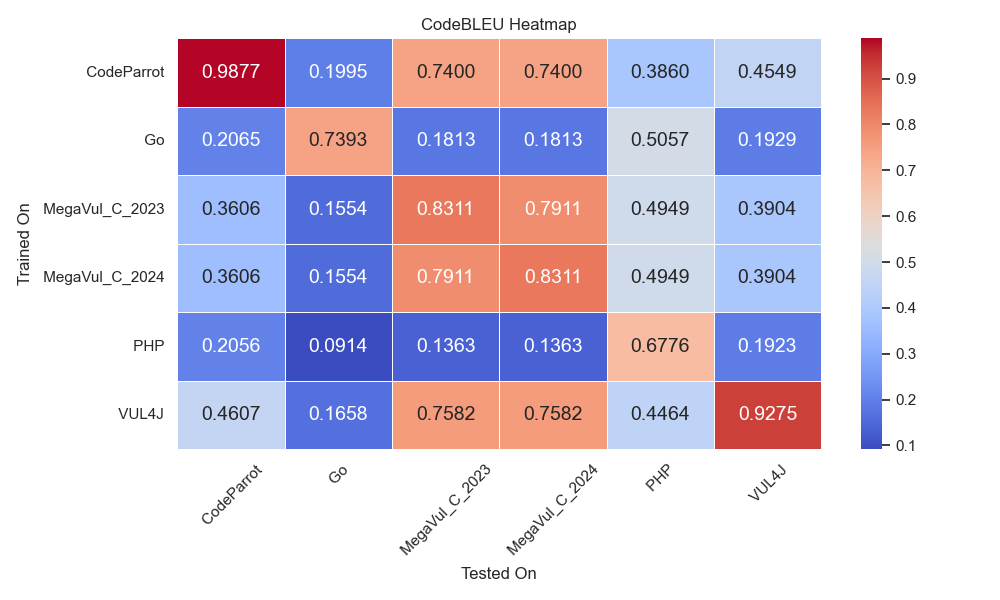}
        \caption{CodeBLEU for CodeT5}
        \label{fig:codet5_codebleu}
    \end{subfigure}
    \hspace{0.02\textwidth}
    \begin{subfigure}[b]{\textwidth}
        \includegraphics[width=\textwidth]{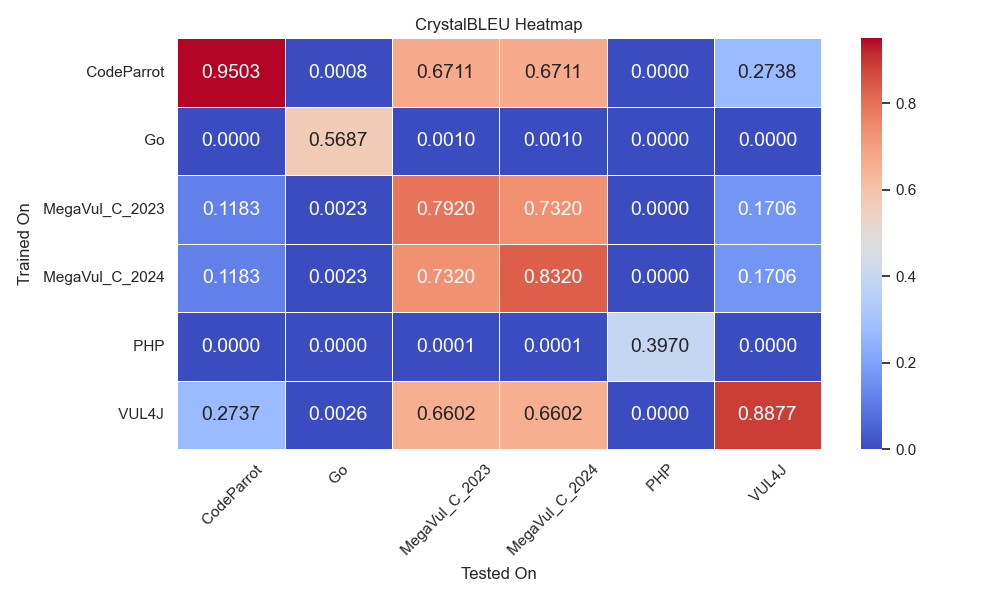}
        \caption{CrystalBLEU for CodeT5}
        \label{fig:codet5_crystalbleu}
    \end{subfigure}
    \captionsetup{justification=centering} 
    \caption{Heatmaps for CodeT5}
    \label{fig2:codet5}
\vspace{-1.5em}
\end{figure}

\section{Results}\label{sec:results}

In this section, we present our findings, focusing on how effectively CodeBERT and CodeT5 generate accurate patches for both known and unknown vulnerabilities across diverse datasets.

\subsection{RQ1: How effectively do CodeBERT and CodeT5 generate accurate patches for known vulnerabilities across diverse datasets?}\label{accuracy-known-vuln}

In this research question, we evaluated the effectiveness of CodeBERT and CodeT5 in generating patches by fine-tuning them on the same dataset. Our analysis spans six datasets across four programming languages, following the methodology outlined in Section~\ref{subsec:test-train-split}. \textbf{Table~\ref{table:rq1_results}}  displays the CodeBLEU, and CrystalBLEU scores of CodeBERT and CodeT5 across six datasets used in our evaluation. Examining the performance of both models on these datasets reveals key insights into how pre-training data diversity and model architecture impact the models’ effectiveness in vulnerability patching tasks. CodeT5 consistently outperforms CodeBERT 
in VUL4J and CodeParrot datasets, with less difference on MegaVul\_C\_2023 and MegaVul\_C\_2024 datasets 
but still demonstrates a clear advantage over CodeBERT when evaluated using both CodeBLEU and CrystalBLEU accuracy scores. This result aligns with the fact that CodeT5 has been pre-trained on diverse data that spans a variety of programming languages and textual formats, enabling it to capture more generalized patterns and nuances in code. On Go, and PHP, CodeBERT performs better than 
CodeT5 using CodeBLEU and CrystalBLEU metrics. By analyzing 
these two datasets, we observed that they often contain incomplete functions or isolated snippets 
lacking full context. This could potentially lead to lower performance for CodeT5, as it relies 
on contextual understanding from diverse sources that might not align well with fragmented or 
incomplete code. Conversely, CodeBERT, which is also trained on a broad variety of programming 
languages, may still benefit from its fine-tuned focus on code structure, making it more adaptable to such fragments.

These findings suggest that CodeBERT's architecture might be inherently more robust when handling 
incomplete or context-limited code, a factor that could contribute to its better performance on Go 
and PHP. Moreover, despite CodeBERT generally being outperformed by CodeT5, lacking the extensive pre-training diversity of 
CodeT5, can still achieve near-competitive results in certain domains, particularly for language-specific tasks.  This observation underscores the need for further investigation to better 
understand the interplay between dataset characteristics and metric sensitivity, rather 
than drawing generalized conclusions about the performance of CodeBLEU or CrystalBLEU.

Our results highlight the benefits of model diversity in deep learning-based vulnerability 
patching. CodeT5's broad pre-training excels on datasets with complex vulnerabilities, 
while CodeBERT's focused design performs well on datasets with more traditional, 
syntactically constrained samples. These insights show that model choice should 
depend on dataset characteristics. CodeBERT's simpler architecture likely makes 
it less reliant on context, while CodeT5 handles diverse inputs more effectively. 
Thus, while CodeT5 is suited for complex, varied data, CodeBERT is valuable in 
environments with incomplete or non-standard code snippets.

\subsection{RQ2: How effectively do CodeBERT and CodeT5 generate accurate patches for \texttt{unknown} vulnerabilities across diverse datasets?}\label{accuracy-unknown-vuln}

\textbf{Figures~\ref{fig1:codebert} and \ref{fig2:codet5}} show the results for RQ2, where we evaluated the fine-tuned CodeBERT and CodeT5 models. For each model, we fine-tuned them on one dataset and tested their performance on two types of datasets: (1) the same dataset used for fine-tuning (in-distribution testing) and (2) all remaining datasets that were not used during fine-tuning (out-of-distribution testing). This setup allowed us to analyze whether fine-tuning pre-trained models (i.e., CodeBERT and CodeT5) on high-quality datasets enhances their ability to detect vulnerabilities, including previously unknown ones. Specifically, we aimed to determine if fine-tuning improves the models' generalization capabilities compared to their pre-trained versions, both on datasets they were trained on and on unseen datasets.

In Section~\ref{accuracy-known-vuln}, we demonstrated the performance of pre-trained models (CodeBERT and CodeT5) in detecting vulnerabilities accurately on datasets they were trained or fine-tuned on. For RQ2, we extended this analysis to evaluate their performance on both in-distribution and out-of-distribution datasets. From Figure~\ref{fig1:codebert} and Figure~\ref{fig2:codet5}, we observe that both models perform significantly better on in-distribution datasets, with almost similar results and only minor percentage differences compared to their performance in Section~\ref{accuracy-known-vuln} or RQ1. This behavior is expected, as fine-tuning allows models to adapt to the specific characteristics of the training data, leading to higher accuracy on familiar datasets.

However, when tested on out-of-distribution datasets, the models exhibit a noticeable drop in accuracy. Notably, when a model trained on a specific programming language is tested on the same language—for example, trained on Megavul\_C\_2023 and tested on Megavul\_C\_2024—the accuracy remains high. A similar trend is observed for Vul4J and CodeParrot, as both are Java-based datasets.
In contrast, for CodeBERT, training on Vul4J and testing on Vul4J results in lower accuracy. This is due to the same reason mentioned in Section~\ref{accuracy-known-vuln}. These findings suggest that while fine-tuning enhances performance on datasets similar to the training data, it does not generalize well to entirely new datasets.

Additionally, the models exhibit non-deterministic behavior (e.g., small variations in accuracy even on in-distribution datasets), which is common in large language models (LLMs) like CodeBERT and CodeT5. This variability can be attributed to factors such as randomness in weight initialization, optimization processes, or inherent fluctuations in the models' predictions.

\section{Discussion}
Fine-tuning on well-characterized datasets substantially boosts CodeBERT and CodeT5 performance in in-distribution tests. However, this advantage drops sharply on out-of-distribution data, especially when the code differs in language or structure. Such declines reflect overfitting, as models learn dataset-specific signals rather than broader security principles.

Additionally, we observe sporadic variability across executions, caused by random weight initialization and hyperparameter sensitivity. Repeated training can alleviate these fluctuations, but consistent checkpointing and parameter tuning remain critical for stable outcomes.

A key lesson is that diverse datasets foster more generalizable repair models. Narrow data coverage may yield high accuracy for certain vulnerability types but struggles with unseen threats. Beyond standard fine-tuning, future work could explore meta-learning, multi-task strategies, and data augmentation to improve cross-domain robustness and ensure patches address genuine security concerns.

\section{Threats to Validity}\label{sec:threats-to-validity}
\textbf{Construct Validity.}
We evaluate ``correct'' patches using CodeBLEU and CrystalBLEU, which primarily gauge syntactic and limited semantic cues. Although these metrics are well-suited for code-focused tasks, they may overlook deeper security implications and potential exploit vectors. Moreover, the labeled “patched” instances within our datasets may not fully represent truly secure fixes, raising the risk of overestimating model performance.

\textbf{Internal Validity.}
Our findings are sensitive to model randomness (e.g., weight initialization) and hyperparameter settings. Even minor fluctuations in these variables can skew comparative outcomes. Additionally, data preprocessing steps such as token truncation and comment removal may eliminate vital context needed to generate security-relevant patches. These factors, if not uniformly controlled, limit the consistency and interpretability of our experimental results.

\textbf{External Validity.}
While this work involves six datasets in four languages, real-world projects frequently rely on specialized libraries and domain-specific coding styles. The observed performance drop on out-of-distribution datasets highlights limited cross-domain generalizability. To enhance broader applicability, future work should consider more diverse datasets and investigate meta-learning approaches that better capture variability across language ecosystems and security contexts.


\section{Conclusion}
Our findings illustrate the promise of large language models for automated vulnerability repair while underscoring significant generalization challenges. CodeBERT and CodeT5 both excel when confronted with familiar vulnerability patterns, yet exhibit performance gaps on unseen datasets and in cross-language contexts. Achieving robust, production-grade vulnerability repair will demand more than simple fine-tuning; it calls for richer datasets, more advanced training paradigms, and continuous adaptation to evolving security threats. By addressing these gaps, future research and practice can more confidently integrate automated patch generation into real-world software development pipelines.

\section{Data Availability}
In support of Open Science, the source code and datasets used in our study are publicly available on Zenodo~\cite{zenodo_artifact}.

\section*{Acknowledgments}
This research received funding from the European Commission through the Horizon Europe Programme as part of the LAZARUS project (\url{https://lazarus-he.eu/}) (Grant Agreement No. 101070303). The content of this article represents the sole responsibility of the authors and does not necessarily reflect the official views of the European Union.



\newpage
\bibliographystyle{splncs04}
\bibliography{references}

\begin{thebibliography}{10}
\providecommand{\url}[1]{\texttt{#1}}
\providecommand{\urlprefix}{URL }
\providecommand{\doi}[1]{https://doi.org/#1}

\bibitem{AlbaneseCJ14}
Albanese, M., {\c{C}}am, H., Jajodia, S.: Automated cyber situation awareness
  tools and models for improving analyst performance. In: Cybersecurity Systems
  for Human Cognition Augmentation, Advances in Information Security, vol.~61,
  pp. 47--60. Springer (2014). \doi{10.1007/978-3-319-10374-7\_3},
  \url{https://doi.org/10.1007/978-3-319-10374-7\_3}

\bibitem{static2017andrei}
Arusoaie, A., Ciobâca, S., Craciun, V., Gavrilut, D., Lucanu, D.: A comparison
  of open-source static analysis tools for vulnerability detection in c/c++
  code. In: 2017 19th International Symposium on Symbolic and Numeric
  Algorithms for Scientific Computing (SYNASC). pp. 161--168 (2017).
  \doi{10.1109/SYNASC.2017.00035}

\bibitem{BuiPVMS24}
Bui, Q., Paramitha, R., Vu, D., Massacci, F., Scandariato, R.: Apr4vul: an
  empirical study of automatic program repair techniques on real-world java
  vulnerabilities. Empir. Softw. Eng.  \textbf{29}(1), ~18 (2024).
  \doi{10.1007/S10664-023-10415-7},
  \url{https://doi.org/10.1007/s10664-023-10415-7}

\bibitem{vul4j2022}
Bui, Q.C., Scandariato, R., Ferreyra, N.E.D.: Vul4j: A dataset of reproducible
  java vulnerabilities geared towards the study of program repair techniques.
  In: 2022 IEEE/ACM 19th International Conference on Mining Software
  Repositories (MSR). pp. 464--468 (2022). \doi{10.1145/3524842.3528482}

\bibitem{Cadar2008klee}
Cadar, C., Dunbar, D., Engler, D.: Klee: unassisted and automatic generation of
  high-coverage tests for complex systems programs. In: 8th USENIX Conference
  on Operating Systems Design and Implementation. p. 209–224. OSDI'08, USENIX
  Association, USA (2008)

\bibitem{Chakraborty2022edit}
Chakraborty, S., Ding, Y., Allamanis, M., Ray, B.: Codit: code editing with
  tree-based neural models. IEEE Transactions on Software Engineering
  \textbf{48}(4),  1385--1399 (2018). \doi{10.1109/TSE.2020.3020502}

\bibitem{SequenceR2021}
Chen, Z., Kommrusch, S., Tufano, M., Pouchet, L.N., Poshyvanyk, D., Monperrus,
  M.: Sequencer: sequence-to-sequence learning for end-to-end program repair.
  IEEE Transactions on Software Engineering  \textbf{47}(09),  1943--1959
  (September 2021). \doi{10.1109/TSE.2019.2940179}

\bibitem{Dang2024}
Dang, N.N.H., Thanh, T.Q., Nguyen-Duc, A.: BERTVRepair: On the Adoption of
  CodeBERT for Automated Vulnerability Code Repair, pp. 173--196. Springer
  Nature Switzerland, Cham (2024). \doi{10.1007/978-3-031-55642-5_8},
  \url{https://doi.org/10.1007/978-3-031-55642-5_8}

\bibitem{deFiteroDominguezGGM24}
de{-}Fitero{-}Dominguez, D., Garc{\'{\i}}a{-}L{\'{o}}pez, E.,
  Garc{\'{\i}}a{-}Cabot, A., Mart{\'{\i}}nez{-}Herr{\'{a}}iz, J.J.: Enhanced
  automated code vulnerability repair using large language models. Eng. Appl.
  Artif. Intell.  \textbf{138},  109291 (2024).
  \doi{10.1016/J.ENGAPPAI.2024.109291},
  \url{https://doi.org/10.1016/j.engappai.2024.109291}

\bibitem{Ding2020metaphor}
Ding, Y., Ray, B., Devanbu, P., Hellendoorn, V.J.: Patching as translation: the
  data and the metaphor. In: Proceedings of the 35th IEEE/ACM International
  Conference on Automated Software Engineering. p. 275–286. ASE'20,
  Association for Computing Machinery, New York, NY, USA (2021).
  \doi{10.1145/3324884.3416587}, \url{https://doi.org/10.1145/3324884.3416587}

\bibitem{eghbali2022crystalbleu}
Eghbali, A., Pradel, M.: Crystalbleu: precisely and efficiently measuring the
  similarity of code. In: Proceedings of the 37th IEEE/ACM International
  Conference on Automated Software Engineering. pp. 1--12 (2022)

\bibitem{feng2020codebert}
Feng, Z., Guo, D., Tang, D., Duan, N., Feng, X., Gong, M., Shou, L., Qin, B.,
  Liu, T., Jiang, D., et~al.: Codebert: A pre-trained model for programming and
  natural languages. arXiv preprint arXiv:2002.08155  (2020)

\bibitem{codebert2020}
Feng, Z., Guo, D., Tang, D., et~al.: Codebert: a pre-trained model for
  programming and natural languages. In: Findings of the Association for
  Computational Linguistics: EMNLP 2020. pp. 1536--1547. Association for
  Computational Linguistics (November 2020).
  \doi{10.18653/v1/2020.findings-emnlp.139}

\bibitem{abs-2012-11701}
Garg, A., Degiovanni, R., Jimenez, M., Cordy, M., Papadakis, M., Traon, Y.L.:
  Learning to predict vulnerabilities from vulnerability-fixes: {A} machine
  translation approach. CoRR  \textbf{abs/2012.11701} (2020),
  \url{https://arxiv.org/abs/2012.11701}

\bibitem{GargDJCPT22}
Garg, A., Degiovanni, R., Jimenez, M., Cordy, M., Papadakis, M., Traon, Y.L.:
  Learning from what we know: How to perform vulnerability prediction using
  noisy historical data. Empir. Softw. Eng.  \textbf{27}(7), ~169 (2022).
  \doi{10.1007/S10664-022-10197-4},
  \url{https://doi.org/10.1007/s10664-022-10197-4}

\bibitem{abs-2303-04247}
Garg, A., Degiovanni, R., Papadakis, M., Traon, Y.L.: Vulnerability mimicking
  mutants. CoRR  \textbf{abs/2303.04247} (2023).
  \doi{10.48550/ARXIV.2303.04247},
  \url{https://doi.org/10.48550/arXiv.2303.04247}

\bibitem{GargDPT24}
Garg, A., Degiovanni, R., Papadakis, M., Traon, Y.L.: On the coupling between
  vulnerabilities and llm-generated mutants: {A} study on vul4j dataset. In:
  {IEEE} Conference on Software Testing, Verification and Validation, {ICST}
  2024, Toronto, ON, Canada, May 27-31, 2024. pp. 305--316. {IEEE} (2024).
  \doi{10.1109/ICST60714.2024.00035},
  \url{https://doi.org/10.1109/ICST60714.2024.00035}

\bibitem{GargPKT24}
Garg, A., Patsakis, C., Khan, Z.A., Tang, Q.: Payload analysis of
  adversaries’ tooling: Automated identification of fuzzers. techrxiv
  preprint  (Dec 2024). \doi{10.36227/techrxiv.173385946.65994728/v1},
  \url{http://dx.doi.org/10.36227/techrxiv.173385946.65994728/v1}

\bibitem{GharibiSF24}
Gharibi, R., Sadreddini, M.H., Fakhrahmad, S.M.: {T5APR:} empowering automated
  program repair across languages through checkpoint ensemble. J. Syst. Softw.
  \textbf{214},  112083 (2024). \doi{10.1016/J.JSS.2024.112083},
  \url{https://doi.org/10.1016/j.jss.2024.112083}

\bibitem{goues2019automated}
Goues, C.L., Pradel, M., Roychoudhury, A.: Automated program repair.
  Communications of the ACM  \textbf{62}(12),  56--65 (2019).
  \doi{10.1145/3318162}

\bibitem{GuoRLFT0ZDSFTDC21}
Guo, D., Ren, S., Lu, S., Feng, Z., Tang, D., Liu, S., Zhou, L., Duan, N.,
  Svyatkovskiy, A., Fu, S., Tufano, M., Deng, S.K., Clement, C.B., Drain, D.,
  Sundaresan, N., Yin, J., Jiang, D., Zhou, M.: Graphcodebert: Pre-training
  code representations with data flow. In: 9th International Conference on
  Learning Representations, {ICLR} 2021, Virtual Event, Austria, May 3-7, 2021.
  OpenReview.net (2021), \url{https://openreview.net/forum?id=jLoC4ez43PZ}

\bibitem{guo2023empirical}
Guo, Y., Hu, Q., Tang, Q., Traon, Y.L.: An empirical study of the imbalance
  issue in software vulnerability detection. In: European Symposium on Research
  in Computer Security. pp. 371--390. Springer (2023)

\bibitem{KhanP18}
Khan, S., Parkinson, S.: Review into state of the art of vulnerability
  assessment using artificial intelligence. In: Guide to Vulnerability Analysis
  for Computer Networks and Systems - An Artificial Intelligence Approach, pp.
  3--32. Computer Communications and Networks, Springer (2018).
  \doi{10.1007/978-3-319-92624-7\_1},
  \url{https://doi.org/10.1007/978-3-319-92624-7\_1}

\bibitem{zenodo_artifact}
Khan, Z.A., Garg, A., Tang, Q.: Artifact for a multi-dataset evaluation of
  models for automated vulnerability repair (2025).
  \doi{10.5281/zenodo.15599983}

\bibitem{Khan0BB22}
Khan, Z.A., Shin, D., Bianculli, D., Briand, L.C.: Guidelines for assessing the
  accuracy of log message template identification techniques. In: 44th
  {IEEE/ACM} 44th International Conference on Software Engineering, {ICSE}
  2022, Pittsburgh, PA, USA, May 25-27, 2022. pp. 1095--1106. {ACM} (2022).
  \doi{10.1145/3510003.3510101}, \url{https://doi.org/10.1145/3510003.3510101}

\bibitem{KhanSBB24}
Khan, Z.A., Shin, D., Bianculli, D., Briand, L.C.: Impact of log parsing on
  deep learning-based anomaly detection. Empir. Softw. Eng.  \textbf{29}(6),
  ~139 (2024). \doi{10.1007/S10664-024-10533-W},
  \url{https://doi.org/10.1007/s10664-024-10533-w}

\bibitem{Kim2013automatic}
Kim, D., Nam, J., Song, J., Kim, S.: Automatic patch generation learned from
  human-written patches. In: Proceedings of the 2013 International Conference
  on Software Engineering. p. 802–811. ICSE '13, IEEE (2013).
  \doi{10.1109/ICSE.2013.6606626}

\bibitem{Le2012genprog}
Le~Goues, C., Dewey-Vogt, M., Forrest, S., Weimer, W.: A systematic study of
  automated program repair: fixing 55 out of 105 bugs for \$8 each. In: 34th
  International Conference on Software Engineering (ICSE). pp. 3--13 (2012).
  \doi{10.1109/ICSE.2012.6227211}

\bibitem{liu2019tbar}
Liu, K., Koyuncu, A., Kim, D., Bissyand\'{e}, T.F.: Tbar: revisiting
  template-based automated program repair. In: Proceedings of the 28th ACM
  SIGSOFT International Symposium on Software Testing and Analysis. p. 31–42.
  ISSTA 2019, Association for Computing Machinery, New York, NY, USA (2019).
  \doi{10.1145/3293882.3330577},
  \url{https://doi-org.proxy.bnl.lu/10.1145/3293882.3330577}

\bibitem{angelix2016}
Mechtaev, S., Yi, J., Roychoudhury, A.: Angelix: scalable multiline program
  patch synthesis via symbolic analysis. In: 38th International Conference on
  Software Engineering (ICSE). pp. 691--701 (2016).
  \doi{10.1145/2884781.2884807}

\bibitem{semfix2013}
Nguyen, H.D.T., Qi, D., Roychoudhury, A., Chandra, S.: Semfix: program repair
  via semantic analysis. In: 35th International Conference on Software
  Engineering (ICSE). pp. 772--781 (2013). \doi{10.1109/ICSE.2013.6606623}

\bibitem{10.1145/3643991.3644886}
Ni, C., Shen, L., Yang, X., Zhu, Y., Wang, S.: Megavul: A c/c++ vulnerability
  dataset with comprehensive code representations. In: Proceedings of the 21st
  International Conference on Mining Software Repositories. MSR '24,
  Association for Computing Machinery, New York, NY, USA (2024).
  \doi{10.1145/3643991.3644886}, \url{https://doi.org/10.1145/3643991.3644886}

\bibitem{ogata2018vetting}
Ogata, M., Franklin, J., Voas, J., Sritapan, V., Quirolgico, S.: Vetting the
  security of mobile applications. Tech. rep., National Institute of Standards
  and Technology (2019). \doi{10.6028/NIST.SP.800-163r1}

\bibitem{OkutanMMKSGS23}
Okutan, A., Mell, P., Mirakhorli, M., Khokhlov, I., Santos, J.C.S., Gonzalez,
  D., Simmons, S.: Empirical validation of automated vulnerability curation and
  characterization. {IEEE} Trans. Software Eng.  \textbf{49}(5),  3241--3260
  (2023). \doi{10.1109/TSE.2023.3250479},
  \url{https://doi.org/10.1109/TSE.2023.3250479}

\bibitem{papineni2002bleu}
Papineni, K., Roukos, S., Ward, T., Zhu, W.J.: Bleu: a method for automatic
  evaluation of machine translation. In: Proceedings of the 40th annual meeting
  of the Association for Computational Linguistics. pp. 311--318 (2002)

\bibitem{PiskachevBB23}
Piskachev, G., Becker, M., Bodden, E.: Can the configuration of static analyses
  make resolving security vulnerabilities more effective? - {A} user study.
  Empir. Softw. Eng.  \textbf{28}(5), ~118 (2023).
  \doi{10.1007/S10664-023-10354-3},
  \url{https://doi.org/10.1007/s10664-023-10354-3}

\bibitem{ren2020codebleu}
Ren, S., Guo, D., Lu, S., Zhou, L., Liu, S., Tang, D., Sundaresan, N., Zhou,
  M., Blanco, A., Ma, S.: Codebleu: a method for automatic evaluation of code
  synthesis. arXiv preprint arXiv:2009.10297  (2020)

\bibitem{RisseB24}
Risse, N., B{\"{o}}hme, M.: Uncovering the limits of machine learning for
  automatic vulnerability detection. In: 33rd {USENIX} Security Symposium,
  {USENIX} Security 2024, Philadelphia, PA, USA, August 14-16, 2024. {USENIX}
  Association (2024),
  \url{https://www.usenix.org/conference/usenixsecurity24/presentation/risse}

\bibitem{wang2021codet5}
Wang, Y., Wang, W., Joty, S., Hoi, S.C.: Codet5: Identifier-aware unified
  pre-trained encoder-decoder models for code understanding and generation.
  arXiv preprint arXiv:2109.00859  (2021)

\bibitem{WuJPLD0BS23}
Wu, Y., Jiang, N., Pham, H.V., Lutellier, T., Davis, J., Tan, L., Babkin, P.,
  Shah, S.: How effective are neural networks for fixing security
  vulnerabilities. In: Proceedings of the 32nd {ACM} {SIGSOFT} International
  Symposium on Software Testing and Analysis, {ISSTA} 2023, Seattle, WA, USA,
  July 17-21, 2023. pp. 1282--1294. {ACM} (2023).
  \doi{10.1145/3597926.3598135}, \url{https://doi.org/10.1145/3597926.3598135}

\bibitem{Nopol2017}
Xuan, J., Martinez, M., DeMarco, F., Clément, M., Marcote, S.L., Durieux, T.,
  Le~Berre, D., Monperrus, M.: Nopol: automatic repair of conditional statement
  bugs in java programs. IEEE Transactions on Software Engineering
  \textbf{43},  34--55 (2017). \doi{10.1109/TSE.2016.2560811}

\bibitem{zhao2020maritime}
Zhao, R., Wang, J., Zheng, X., Wen, J., Rao, L., Zhao, J.: Maritime visible
  image classification based on double transfer method. IEEE Access
  \textbf{8},  166335--166346 (2020)

\end{thebibliography}

\end{document}